\documentclass[12pt]{article}
\usepackage{graphicx}

\begin{document}

\bigskip

\noindent
{\bf Students Watching Stars Evolve}

\bigskip

\noindent
{\bf John Percy, Drew MacNeil, Leila Meema-Coleman, and Karen Morenz}, Department of Astronomy and Astrophysics, University of Toronto, Toronto ON, Canada

\bigskip
\noindent
February, 2012

\bigskip

\noindent
{\bf Abstract}
\smallskip
We describe a study of period changes in 59 RR Lyrae stars, using times
of maximum brightness from the GEOS database.  The work was carried out
by outstanding senior high school students in the University of Toronto
Mentorship Program.  This paper is written in such a way that high school
or undergraduate physics and astronomy students could use it as a guide and
template for carrying out original research, by studying period changes
in these and other types of pulsating variable stars.

\bigskip

\noindent
{\bf Introduction}

\smallskip

The sun is a star, one of hundreds of billions in our Milky Way galaxy,
which is one of tens of billions in the universe.  Our sun makes life
possible on Earth, and other stars may nourish life on their own planets.
As stars radiate energy, however, they gradually exhaust their fuel, leading to
change and eventual death.

A star is an immense sphere of gas, about 3/4 hydrogen, 1/4 helium, and
2\% heavier elements, typically a million times larger and more massive
than Earth.  Throughout most of its life, it generates energy in
its hot, dense core by thermonuclear fusion of hydrogen into helium, currently at
the rate of 4 x $10^{26}$ W, in the case of the sun.
Gradually the hydrogen in the core is depleted.
The structure and properties of the star change.  It expands
into a red giant star.  The aging of the star takes billions of years,
but it has direct consequences that can be observed, measured, and analyzed
by students!

\medskip

\noindent
{\bf The Physics of Stars}

\smallskip

The structure of a star is governed by simple laws of physics.  Deeper in
the star, the pressure of the gas is greater, because of
the greater weight of the overlying layers.  Since the gas
generally obeys the perfect gas law, the temperature and density also
increase inward.  The {\it outward} thermal gas pressure gradient balances the
{\it inward} pull of gravity on the gas.  

Heat always flows from hot to less
hot, so energy must flow outwards, either by radiation or convection,
depending on the transparency of the gas to radiation.  If the star is to remain
stable, the outward-flowing
energy must then be replenished by energy generation in the core; otherwise
the core will cool, pressure will decrease, and the star will shrink.  These principles, along with conservation of mass and energy, can be
expressed as four equations.  These, plus the input mass and composition,
the gas law, and information about the transparency of the gas, and about the
thermonuclear energy generation, can be solved on a computer to produce a
"model" of the star -- its physical properties from center to surface.  The changes
in these properties due to thermonuclear fusion of hydrogen into helium
can then be incorporated into the calculations to produce an "evolutionary sequence of models".
These predict how the star will age with time.  They can be tested in many ways.

\medskip
\noindent
{\bf Pulsating Stars}

\smallskip

At various stages in a star's life, its properties make it unstable to
{\it pulsation} -- in-and-out (radial) vibration being the simplest
kind.  The outflowing radiation interacts
with the gas in the outer layers of the star, converting radiant energy
into mechanical energy.  The period P, the time for one vibration, depends
primarily on the radius R of the star, just as the period of vibration of a
string (such as a guitar string) depends primarily on its length.  Specifically,
P $\sim$ $R^{1.5}$.  
As a result of the pulsation,
the star brightens and fades from maximum to minimum brightness and back every
P days.  It is a {\it variable star$^{1,2}$}.

A graph of the brightness of the star versus time is called a {\it light curve}
(Fig. 1).  It plots the "apparent magnitude", which is proportional
to -2.5log(brightness), versus time (usually the "Julian Date"$^{3}$) or
the phase, which is the fraction of a cycle elapsed since maximum brightness.

\begin{figure}
\centering
\includegraphics[height=8cm]{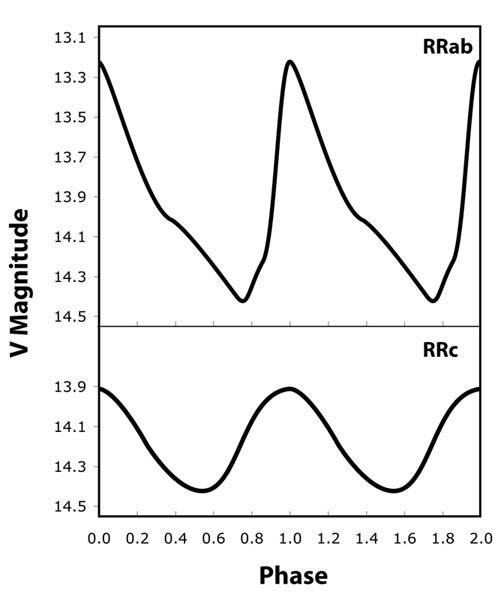}
\caption{Light curves of an RRab (top) and an RRc (bottom) star -- apparent
magnitude (brightness increasing vertically) versus time, in units of cycles
measured from maximum brightness.  RRab stars are pulsating in the fundamental mode,
and have sharp maxima.  RRc stars are pulsating in the first overtone mode,
and have rounded maxima.}
\end{figure}

\begin{figure}
\centering
\includegraphics[height=8cm]{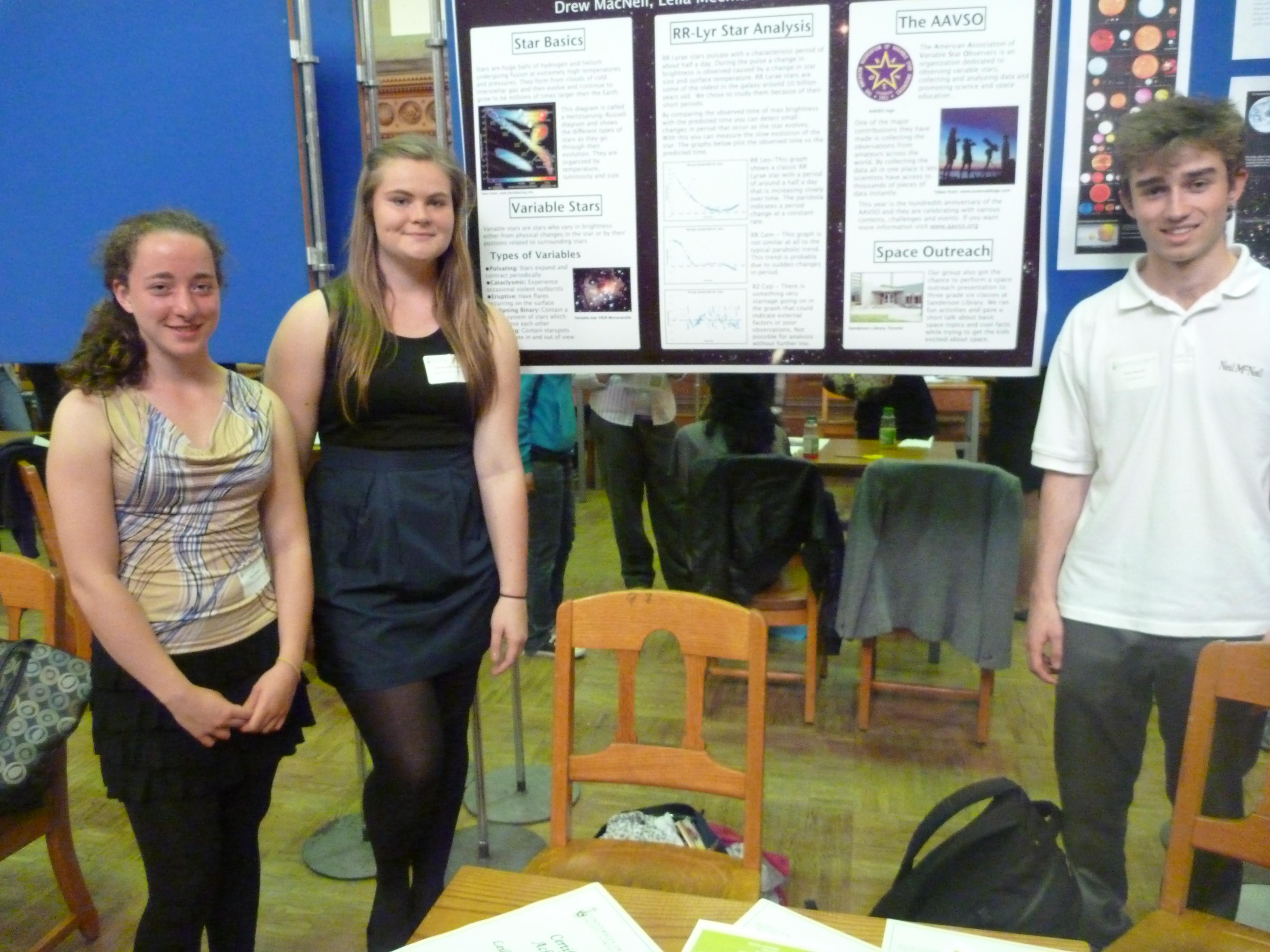}
\caption{Coauthors Morenz, Meema-Coleman, and MacNeil with their poster at
the UTMP reception and research fair, attended by mentors and mentees,
family and teachers.}
\end{figure}

\medskip

\noindent
{\bf Period Changes in Pulsating Stars}

\smallskip

As a pulsating star ages, it initially expands slowly, so the period
lengthens and the separation of the times of maximum brightness 
gradually increases.  [Later in its life, the star may contract, in which
case the period would decrease.]  Specifically: (dP/dt)/P $\sim$ 1.5(dR/dt)/R.  This process of {\it period change}, though slow, is observable because
its effects are {\it cumulative}.  Consider the analogy of a perfect clock A,
and a less perfect clock B that runs one second more slowly every day, starting
with the day when it runs at the correct rate.
The cumulative error in clock B, after 1, 2, 3, 4 and 5 days, relative
to clock A, is: 0, 1,
1+2=3, 3+3=6, 6+4=10 seconds etc.  The cumulative error increases as
the {\it square} of the elapsed time.  The possibility of watching stars
slowly age by measuring the change in their pulsation period was recognized
almost a century ago.

Cepheids, named after the star delta Cephei$^{4}$, are an important class
of pulsating stars.  Their periods, which range from 1 to 100 days,
are correlated with their average power, so their distance can be estimated
from the inverse-square law of brightness by comparing their average
apparent brightness with the power deduced from their pulsation periods.
Polaris (the North Star) is the best-known star in the northern sky.
 It is a Cepheid with a period of 3.96925... days, which is increasing by 4.5 seconds per year as the star ages.

\medskip

\noindent
{\bf Educational Context}

\smallskip

Over the decades, co-author Percy's research program on variable
stars and stellar evolution has been enriched by the participation of
undergraduates, and of outstanding senior high school students through the
University of Toronto Mentorship Program (UTMP).  These students develop
and integrate their science, math, and computing skills, motivated by
doing real science, with real data$^{5}$.  Co-authors MacNeil, Meema-Coleman,
and Morenz were participants in the UTMP in 2011.  They met with Percy regularly;
they produced a poster paper, on this project, for the end-of-year
Research Fair and Reception.  MacNeil and Meema-Coleman also developed and gave an
engaging astronomy outreach program for 70 grade six students at a local
inner-city public library.

\medskip
\noindent
{\bf This Project}

\smallskip

In 2011, for the UTMP, we decided to use pulsation period changes to study the evolution of
RR Lyrae stars$^{6,7}$ (the nomenclature of variable stars, leading to names
such as RR Lyrae, is explained on an excellent ``stars" website$^{8}$ maintained
by Professor Jim Kaler).
They are old sunlike stars, with pulsation periods of about half a day, which
have exhausted the hydrogen fuel in their core, swelled into red giants,
lost about half their mass, and are now fusing helium
into carbon.  Since helium fusion is less efficient than hydrogen fusion,
and because the star is now emitting about 100 times the power of the
sun, it fuels the star for only a few million years, before the star
again exhausts its fuel and again swells into a red giant.

There are complications to be expected.  Some RR Lyrae stars (designated RRab) pulsate in
the fundamental (simplest) mode, others (designated RRc) in the first overtone, and a few (designated RRd)
in both.  Some RR Lyrae stars also show the {\it Blazhko effect$^{9}$}, a
long-term (weeks) variation in the amplitude and shape of the light curve
which may be due to interference between the in-and-out pulsation mode,
and a more complicated ``non-radial" mode of pulsation.

\medskip
\noindent
{\bf Data}

\smallskip

The brightness changes in RR Lyrae stars have been measured for many
decades, especially by skilled amateur astronomers.  The American
Association of Variable Star Observers$^{2}$ is the world leader
in coordinating such measurements; it celebrated its centennial in 2011.
Amateur astronomers and students can make these measurements with a
small telescope and the unaided eye, though most of the measurements
are now made with CCD (charge-coupled device) digital cameras.
We used times of maximum brightness in
a database$^{10}$ maintained by the Groupe Europ\'{e}en d'Observation Stellaire (GEOS), including times from previous publications, from the AAVSO and other
such organizations, and from robotic telescopes$^{11}$.  The database
includes about 50,000 times of maximum of over 3,000 stars, covering up
to a century.  There were 59 stars with data that were long and continuous enough for
straight-forward analysis.

\medskip
\noindent
{\bf Analysis}

\smallskip

For any star, the observed times of maximum brightness t(i) are compared
with the predicted times t(0) + NP where t(0) is an initial time of
maximum, P is the assumed period in days, and N is an integer -- the
number of cycles between t(0) and t(i).  t(i) - (t(0) + NP) is called
{\it (O-C), observed minus calculated}.  A graph of (O-C) versus t -- an (O-C)
diagram or OCD -- will be a
straight line if the period P is constant; its slope will be positive,
zero, or negative, depending on whether P is greater than, equal to,
or less than the true period.  This slope can therefore be used to correct or refine the
existing period of the star.  If P is changing linearly (i.e. P = P(0)
+ kt where k is a constant), then the OCD will be a parabola, opening
upward or downward, depending on whether k is positive or negative.
A parabola can be fitted to the OCD using least-squares
software, such as the fitting function in spreadsheet software such as Excel.

A complication occurs if there are long gaps between 
the t(i): the number of cycles in the gap is uncertain.
If the gaps are not too long, and if the measurements before and after
the gaps are reasonably continuous, it should be possible to infer the
number of cycles in the gap by extrapolation of the existing measurements.

The database was scanned for stars having a sufficient number and
continuity of t(i).  Periods and t(i)s were entered into a spreadsheet;
values of C and of (O-C) were determined; OCDs were plotted.  If their
slopes were non-zero, the existing periods were modified.

We initially classified the OCDs as approximately (i) parabolic (P); (ii) linear (L);
(iii) wavy (W); (iv) otherwise peculiar (?) -- not parabolic or linear; or (v) too
scattered (S) to classify.  We later decided that parabolic-linear was a
continuum in the sense that ``apparently linear" corresponded to a parabola
with very small curvature, and that ``wavy" was a special case of ``peculiar".

The accuracy of the t(i) is greater for the RRab stars, because they have
sharp-peaked light curves, whereas the RRc have more sinusoidal ones (Fig. 1).
The accuracy is generally also greater if the t(i) has been measured with a CCD camera,
rather than with the unaided eye or photography.  The accuracies are typically a few minutes.

A characteristic time scale, for period change and therefore for change in
the size of the star, is the period divided by the
rate of change of the period P/(dP/dt), in appropriate units, i.e. it is
the time required for the period to change by P.  We designate the time
scale by $\tau$ (tau).  This is approximately the length of time
that the RR Lyrae phase of evolution would last, though that time more
correctly depends on how long the star's properties cause it to be unstable
to pulsation.  Alternatively, the
time scale could be defined as the time required for the period to change by a factor
of 2, or {\it e}.  These time scales apply to the few decades which are covered
by our data, though evolutionary models predict that the rate of period change
should be uniform over millions of years.

\medskip
\noindent
{\bf Results}

\smallskip

Tables 1 and 2 list the stars analyzed: their name, period (day), (O-C)
diagram type, RR Lyrae subtype, [Fe/H], dP/dt in seconds per 100 years, and the characteristic
evolution time $\tau$ = P/dP/dt in millions of years.  [Fe/H] is the logarithm of the
ratio of a star's iron abundance to its hydrogen abundance, compared with
that of the sun; it is used as a measure of the abundance of elements heavier
than helium.

\begin{enumerate}

\item
Of the 59 stars that were analyzed, 10 were too scattered (S) to classify, 27
were parabolic (P), 14 were linear (L), P + L = 41 in total; 3 were wavy (W) and 5 were
otherwise peculiar (?), W + ? = 8 in total.

\item
The two RRd stars were both among the stars with too much scatter for
analysis; their period changes must be studied in more sophisticated ways.

\item
The Blazhko-effect stars showed no particular tendency for scatter, though
5 of the 18 had peculiar OCDs.

\item
Of the RRab stars whose (O-C) diagrams could be fit with parabolas, 15 had
increasing periods (median $\tau$ of 10 million years) and 14 had decreasing
periods (median $\tau$ of 6.7 million years) i.e. there were approximately
the same number increasing as decreasing, and the rates of evolution were
similar.

\item
The RRc stars have sinusoidal light curves.  The times of maximum are more
difficult to measure than for the RRab stars, whose light curves have sharp
maxima, so they were not studied by Le Borgne et al.$^{11}$.  Of the 14 RRc
stars in our sample, 4 had scattered OCDs, 2 had peculiar OCDs, 5 showed
increasing periods (median $\tau$ of about 20 million years), and 3 showed 
decreasing periods with a large scatter in the rate.  The median $\tau$ 
suggests rather slow evolution, compared with the RRab stars.  The sample
size, however, is small.

\end{enumerate}

\begin{figure}
\includegraphics[height=8cm]{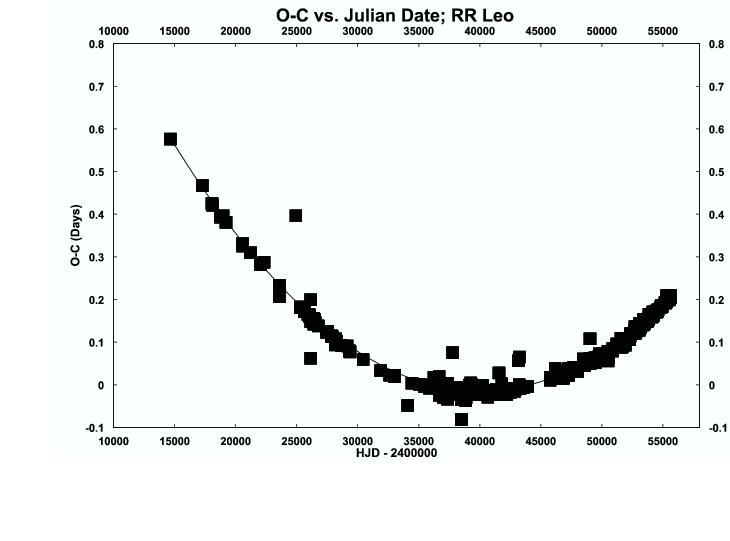}
\caption{The parabolic OCD for RR Leo, corresponding to a linear period
increase of 2.7 seconds per century.}
\end{figure}

Two examples of OCDs are shown in Figs. 3 and 4.  RR Leo has a
parabolic OCD, corresponding to a period increase of 2.7 seconds per
century, and RR Gem has a peculiar one, neither parabolic nor linear.
A few OCDs seem to show abrupt pulsation period changes, which
are difficult to explain theoretically.  Various explanations have been proposed for these peculiar
OCDs, including ``burps" of convection$^{12}$, deep in the star, or 
magnetic cycles in their outer layers. 

\medskip
\noindent
{\bf Discussion}

\smallskip

It is not easy to compare our results in detail with the predictions of evolutionary
models$^{13}$ because the sign and magnitude of the predicted changes depends strongly on the exact mass and
composition of the star, especially the abundance of the elements heavier
than helium (denoted Z).  The value of [Fe/H] (see Tables 1 and 2) is used as
a proxy for Z.  Indeed, the stars in our sample have a wide range
of values of [Fe/H], and probably of masses (though most are probably between 
0.4 and 0.6 times the sun's).  The magnitudes of the observed period 
changes, and the
values of $\tau$ are reasonably consistent with the models, though the models
predict many more period increases than decreases.  See references
11-13 for more detailed comparisons between observations of period changes,
and predictions of evolutionary models.

Pulsation period changes have been studied in other kinds of variable
stars, with a variety of interesting results.
Mira stars are extreme red giant stars which, at the ends of their lives,
pulsate with periods of several hundred days.  In these stars, the pulsation
drives off the outer layers of the star into space, leaving a {\it white
dwarf} corpse, a million times denser than water.  The pulsation of these
stars is complicated by random cycle-to-cycle period fluctuations of a few
percent, leading to wave-like (O-C) diagrams.  Times of maximum brightness
are available for analysis on the AAVSO website$^{14}$. 

\begin{figure}
\centering
\includegraphics[height=8cm]{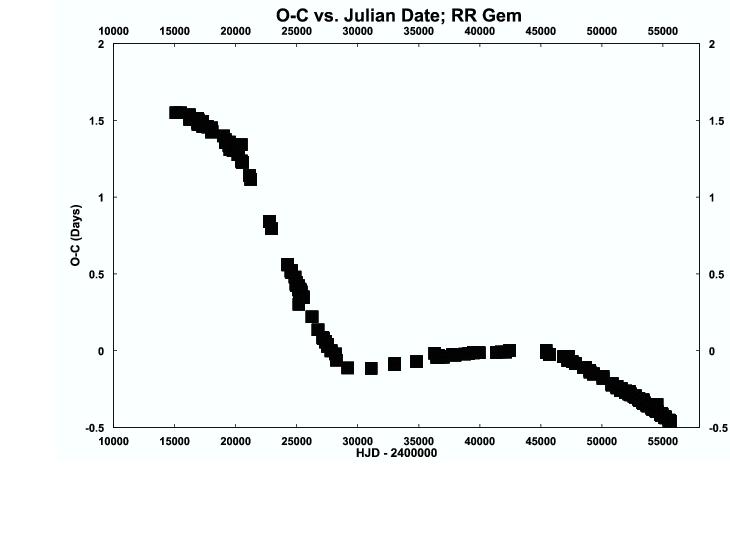}
\caption{The non-parabolic, non-linear OCD for RR Gem.  The OCD is not
explainable as due solely to evolution, but must be due to one or more
other processes.}
\end{figure}






Eclipsing binary stars are pairs of stars in mutual orbit in which one
or both stars periodically eclipses the other star, resulting in a
brief decrease in brightness.  (O-C) analysis of the times of {\it minimum}
brightness, also available on-line$^{15}$ provides information on mass transfer or loss in the system.

The period changes in Cepheid pulsating stars are particularly important
for testing evolutionary models, and have been extensively studied by
Leonid Berdnikov, David Turner, and their colleagues$^{16}$.
High school (and university) students can {\it observe} the variability of Cepheids using the unaided eye$^{17}$ or a
camera$^{18}$, and verify the slow period changes in these stars.  Students'
observations of these or other variable stars can be entered into the AAVSO
International Database, which now contains over 20 million observations, for other researchers to use.

The detectability of linear period changes, due to the aging of the star,
increases as the {\it square} of the timespan of the data, so it is worthwhile
for observers to continue their measurements, and for analysts to repeat 
the (O-C) analysis as longer datasets become available.  For instance: the
period changes in some of the RRab stars in our sample were studied a few
years ago$^{11}$ but the analysis should be repeated periodically.  We
believe that this type of analysis provides an engaging education experience
for undergraduates or for senior high school students; the physical and
mathematical concepts are simple, data is available on-line and, with it,
students can actually observe the effects of the slow aging of the stars.

\medskip
\noindent
{\bf Conclusions}

\smallskip

Students can use existing data on the times of maximum brightness of
pulsating stars to observe and measure the evolutionary changes in
the stars, even though the characteristic times, for the changes, are
millions to billions of years. In this way, they can make meaningful use
of their science and math skills, motivated by doing real science, with
real data.

\medskip
\noindent
{\bf Acknowledgements}

\smallskip
\noindent
We thank the hundreds of observers, mostly skilled amateur astronomers, who
voluntarily made the measurements on which this paper is based, and we thank
GEOS for maintaining their useful database.

\medskip
\noindent
{\bf References}

\smallskip

(1) J.R. Percy, {\it Understanding Variable Stars} (Cambridge University Press, New York, NY, 2007).

(2) www.aavso.org/

(3) www.aavso.org/about-jd

(4) www.aavso.org/vsots\_delcep

(5) J.R. Percy, ``Variable star research projects for outstanding senior high school students," {\it Journal of the AAVSO}, {\bf 35}, 281-283 (2006);
 www.aavso.org/publications/ejaavso/v35n1/284.shtml

(6) www.aavso.org/vsots\_xzcyg

(7) www.aavso.org/vsots\_rrlyr

(8) stars.astro.illinois.edu/sow/starname.html

(9) www.aavso.org/now-less-mysterious-blazhko-effect-rr-lyrae-variables

(10) dbrr.ast.obs-mip.fr

(11) J.F. Le Borgne et al., ``Stellar evolution through the ages: period variations in galactic RRab stars as derived from the GEOS database and TAROT telescopes," {\it Astron. Astrophys.}, {\bf 476}, 307-316 (2007).

(12) A.V. Sweigart and A. Renzini, ``Semiconvection and period changes in RR Lyrae stars," {\it Astron. Astrophys.} {\bf 71}, 66-78 (1979).

(13) Lee, Y.-W., ``Stellar evolution and period changes in RR Lyrae stars," {\it Astrophys. J.} {\bf 367}, 524-527 (1991).

(14) www.aavso.org/maxmin

(15) www.aavso.org/observed-minima-timings-eclipsing-binaries

(16) D.G. Turner et al., ``Rate of period change as a diagnostic of Cepheid properties," {\it Publ. Astron. Soc. Pacific}, {\bf 118}, 410-418 (2006). 

(17) J.R. Percy and A. Rincon, ``Observing Zeta Geminorum as a high school physics project," {\it J. Amer. Assoc. Variable Star Observers}, {\bf 24}, 24-25 (1996); www.aavso.org/publications/ejaavso/v24n1/24.shtml.

(18) J.R. Percy, L. Syczak, and J.A. Mattei, ``Using 35mm slides for measuring variable stars," {\it The Physics Teacher}, {\bf 35}, 349-351 (1997); see also www.aavso.org/publications/ejaavso/v35n1/281.shtml.

\medskip

\noindent
{\bf John Percy} is Professor Emeritus, Astronomy \& Astrophysics, and Science
Education, University of Toronto.  {\bf Drew MacNeil, Leila Meema-Coleman, and
Karen Morenz} were high school students in the prestigious University of
Toronto Mentorship Program in 2011.

\smallskip

\noindent
{\bf Department of Astronomy \& Astrophysics, University of Toronto,
Toronto, ON M5S 3H4, Canada; john.percy@utoronto.ca}

\clearpage

\begin{table}
\begin{tabular}{rrrrrrr}
\hline
Name & Period & OCD & Type & [Fe/H] & dP/dt & $\tau$ \\
 & day & & & & s/100Y & MY \\
\hline
DX Del & 0.47261807 & P & ab & -0.11 & 0.412 & 9.60 \\
EZ Cep & 0.37899919 & P & c & -- & 0.404 & 7.85 \\
EZ Lyr & 0.52526415 & L: & ab & -1.30 & 0.047 & 93.5 \\
GI Gem & 0.43326648 & L: & ab & -- & -0.87 & -4.17 \\
IO Lyr & 0.57712192 & L: & ab & -1.14 & -0.22 & -21.9 \\
OV And & 0.47058076 & L: & ab & -- &-0.116 & -33.9 \\
RR Cet & 0.55302795 & P & ab & -1.60 & 0.260 & 17.8 \\
RR Gem & 0.39731060 & ? & ab & -0.29 & -- & -- \\
RR Leo & 0.45238969 & P & ab & -1.60 & 2.660 & 1.42 \\
RR Lyr & 0.56683295 & L & ab & -1.14 & -0.676 & -7.01 \\
RS Boo & 0.37733834 & P & ab & -1.36 & 0.381 & 8.29 \\
RU CVn & 0.57324490 & W & ab & -- & -- & -- \\
RU Psc & 0.39038500 & S & c & -1.65 & -- & -- \\
RV Cap & 0.44774566 & P & ab & -1.61 & -1.238 & -3.03 \\
RV CrB & 0.33156500 & S & c & -1.69 & -- & -- \\
RV Oct & 0.57116250 & S & ab & -1.71 & -- & -- \\
RV UMa & 0.46806000 & W & ab & -1.04 & -- & -- \\
RW Cnc & 0.54720569 & P: & ab & -1.67 &  1.910 & 2.40 \\
RW Dra & 0.44291700 & ? & ab & -1.55 & -- & -- \\
RX Cet & 0.57371095 & L: & ab & -1.28 & 1.108 & 4.33 \\ 
RZ Cep & 0.30868530 & S & c & -1.77 & -- & -- \\
RZ Lyr & 0.51124591 & P & ab & -1.69 & -1.578 & -2.71 \\
S Com & 0.58658810 & P & ab & -1.91 & -0.892 & -5.50 \\
SS Cnc & 0.36733833 & P & ab & -0.24 & 0.185 & 16.6 \\
SS Tau & 0.36991119 & L & ab & -- & -1.270 & -2.44 \\
ST Boo & 0.62229368 & L & ab & -1.72 & 0.348 & 15.0 \\
ST Leo & 0.47798421 & L & ab & -1.17 & -0.416 & -9.61 \\
SU Dra & 0.66042067 & P & ab & -1.39 & 0.534 & 10.3 \\
SW And & 0.44226806 & P & ab & -0.24 & -1.446 & -2.56 \\
SW Aqr & 0.45930290 & L & ab & -1.63 & 0.067 & 57.4 \\
\hline
\end{tabular}
\caption{Rates of period change, and characteristic evolution times for RR Lyrae stars.}
\end{table}

\clearpage

\begin{table}
\begin{tabular}{rrrrrrr}
\hline
Name & Period & OCD & Type & [Fe/H] & dP/dt & $\tau$ \\
 & day & & & & s/100Y & MY \\
\hline
SW Dra & 0.56967136 & P & ab & -1.12 & 0.353 & 13.5 \\
SX Aqr & 0.53571243 & P & ab & -1.87 & -0.686 & -6.53 \\
V394 Her & 0.43605526 & P: & ab & -- & -0.894 & -4.08 \\
SZ Hya & 0.53724022 & ? & d & -- & -- & -- \\
TU UMa & 0.55765749 & L: & ab & -1.05 & -0.160 & -29.1 \\
SX UMa & 0.30711780 & ? & c & -1.82 & -- & -- \\
AA Aql & 0.36178731 & P: & ab & -0.20 & 0.032 & 94.6 \\
AH Cam & 0.36872857 & B & ab & -- & -4.100 & -0.75 \\
AQ Cnc & 0.54851950 & P: & ab & -- & -0.353 & -13.0 \\
AR Per & 0.42554959 & P & ab & -0.17 & 0.250 & 14.2 \\
AV Peg & 0.39037583 & P & ab & 0.02 & 1.537 & 2.13 \\
AA CMi & 0.47632365 & P & ab & -0.15 & 0.499 & 7.99 \\
AC And & 0.71123760 & S & d & -1.16 & -- & -- \\
AN Ser & 0.52207130 & S & ab & -0.07 & -- & -- \\
AQ Lyr & 0.35714240 & S: & ab & -- & -- & -- \\
AR Her & 0.47002800 & ?: & ab & -1.30 & -- & -- \\
AT And & 0.61691220 & S & ab & -1.18 & -- & -- \\
AR Ser & 0.57514160 & S & ab & -0.85 & -- & -- \\
BC Dra & 0.71958926 & L: & ab & -2.00 & -0.777 & -7.75 \\
BD Dra & 0.58906655 & P: & ab & -- & 4.240 & 1.16 \\
TV Boo & 0.31256050 & P & c & -- & 0.087 & 30.1 \\
AE Boo & 0.31489430 & L: & c & -- & 0.248 & 10.6 \\
ST CVn & 0.32904500 & W & c & -- & -- & -- \\
BB CMi & 0.39642400 & S & c & -- & -- & -- \\
VZ Dra & 0.32102829 & P & c & -- & -1.441 & -1.93 \\
LS Her & 0.23080777 & P: & c & -- & 0.052 & 42.4 \\
TV Lyn & 0.24065125 & P: & c & -- & 0.091 & 22.9 \\
DH Peg & 0.25551056 & L: & c & -- & -0.037 & -60.3 \\
SS Psc & 0.28779248 & P & c & -- & -0.207 & -10.2 \\
\hline
\end{tabular}
\caption{Rates of period change, and characteristic evolution times for RR Lyrae stars (continued).}
\end{table}

\end{document}